\documentclass[a4paper,11pt]{article}
\usepackage{pos}
\usepackage{aas_macros}

\title{Development of PANOSETI Telescopes for Ultra-High-Energy Gamma-Ray Astronomy }

\author*[a]{N. Korzoun}
\emailAdd{nkorzoun@udel.edu}
\onbehalf{for the PANOSETI Collaboration}

\affiliation[a]{
    Department of Physics and Astronomy and the Bartol Research Institute, University of Delaware,\\
    104 The Green, Newark, DE 19716, USA
}

\abstract{
Ultra-High-Energy (UHE, E $>100$\,TeV) gamma rays are one of the few channels to search for and study Galactic PeVatrons. Among the most promising PeVatron candidates are the many UHE gamma-ray sources that have recently been identified on the Galactic Plane. Ground-based particle detectors see these sources as extended rather than point-like, and current generation Imaging Atmospheric Cherenkov Telescopes (IACTs) struggle to study them with effective areas and background rejection that are suboptimal at UHE. A cost-efficient way of constructing an array of IACTs explicitly designed for UHE sensitivity is to sparsely separate many small telescopes. We have simulated, prototyped, and twice deployed a pathfinder array that is instrumented with telescopes designed by the Panoramic Search for Extraterrestrial Intelligence (PANOSETI) team. These 0.5-meter Fresnel lens telescopes are purpose-built for imaging optical transients on nanosecond timescales and are equipped with a $10^\circ\times10^\circ$ silicon photomultiplier camera. Three PANOSETI telescopes were deployed twice in the same temporary configuration at Lick Observatory in March and October 2024. Here we give a brief description of the instrument and present a comparison of simulations with the data collected, including an analysis of the Crab Nebula. We also report on the ongoing deployment of PANOSETI telescopes for the Dark100 array that is planned to operate for five years at Palomar Observatory.
}

\ConferenceLogo{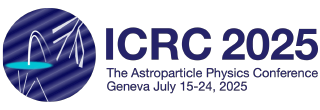}

\FullConference{39th International Cosmic Ray Conference (ICRC2025)\\
 15–24 July 2025\\
Geneva, Switzerland\\}

\begin{document}
\maketitle

\section{Introduction}
The 1LHAASO catalog of gamma-ray sources revitalized UHE gamma-ray astronomy with its announcement that 43 sources on the Galactic Plane emit gamma rays with energies in excess of 100 TeV \cite{2024ApJS..271...25C}. 32 of these object have no known TeV counterpart, so the current generation of IACTs are not optimized at the required energies for followup studies of these mysterious objects. One way of building a new gamma-ray observatory with an angular resolution comparable to traditional IACTs, and an effective area on the same order of magnitude as extensive air shower arrays, is to widely disperse PANOSETI telescopes. Such an array would also be ideal for detailed study of gamma-ray emission near the Galactic Center \cite{2016Natur.531..476H, 2024ApJ...973L..34A}, and for searching for signatures of ultra-heavy dark matter.
 
\begin{figure}[htb]
    \centering
    \includegraphics[width=\linewidth]{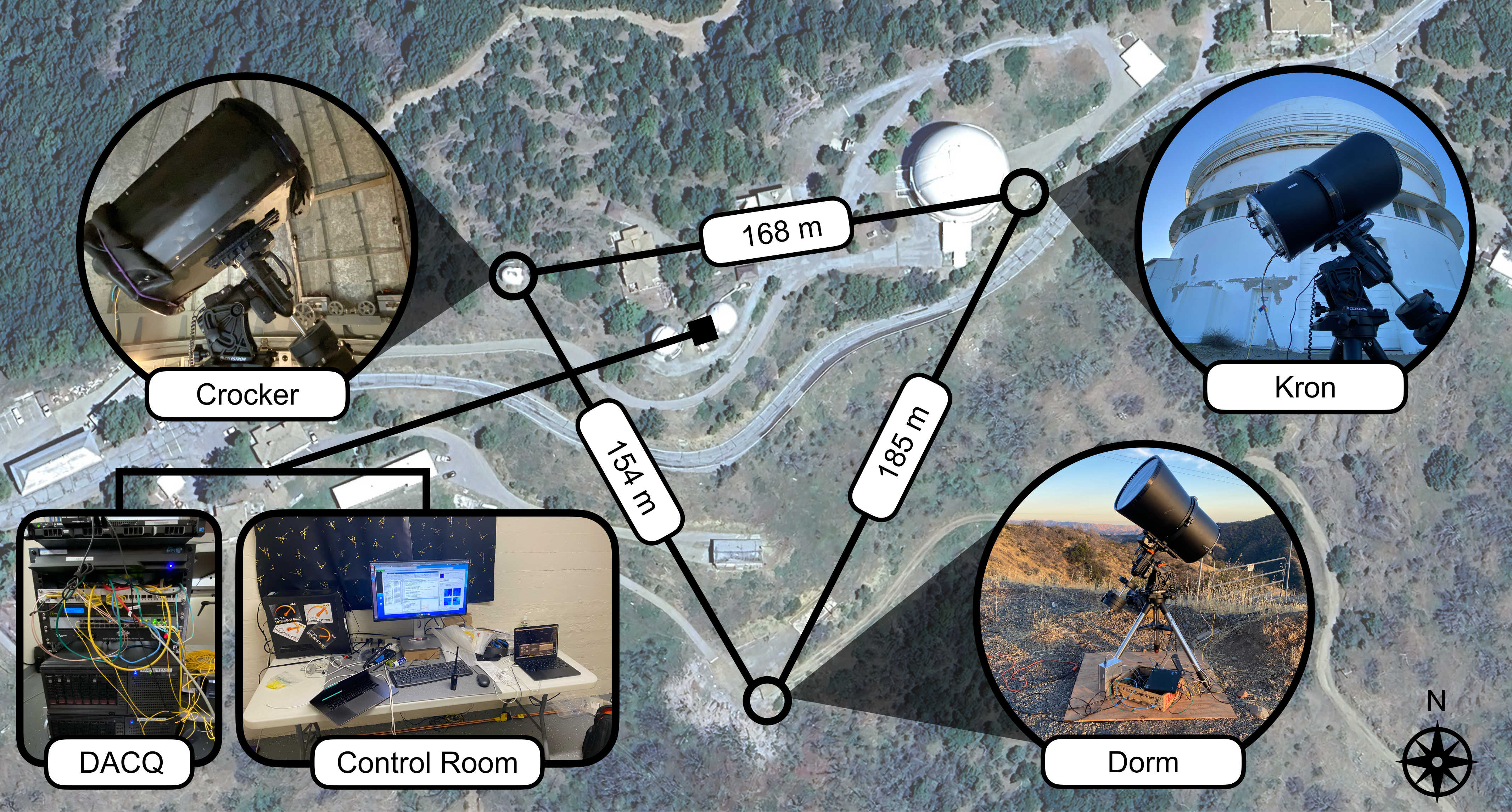}
    \caption{Layout of the 2024 PANOSETI deployments at Lick Observatory. Crocker, Kron, and Dorm are the names of the sites the telescopes were deployed at. All telescopes were operated with a single computer in the control room, and the data streams were collected by the same data acquisition unit (DACQ). Maps Data: Google, \copyright 2025 Airbus.}
    \label{fig:lick}
\end{figure}

PANOSETI telescopes were originally designed for the purpose of searching for optical transients on nanosecond timescales \cite{2018SPIE10702E..5IW}. Their large $10^\circ\times10^\circ$ field of view, 1024 pixel silicon photomultiplier camera, and small 0.5-meter Fresnel lens aperture are all design elements that incidentally make PANOSETI telescopes excellent devices for imaging the air showers created by UHE gamma rays. In 2021, two PANOSETI telescopes were used to conduct simultaneous observations of the Crab Nebula with VERITAS \cite{2022SPIE12184E..8BM}. The results showed PANOSETI telescopes have the potential to be operated as an IACT that is capable of imaging air showers produced by UHE gamma rays. This motivated preliminary simulations which determined the PANOSETI energy threshold to be tens of TeV \cite{2023arXiv230809607K}. Subsequently, a pathfinder array was twice deployed at Lick Observatory in the same configuration -- a triangle with average telescope spacing of 169\,m (Figure \ref{fig:lick}). The dataset from the first deployment was primarily used to compare simulations with real data so that a rudimentary set of analysis tools could be developed for analyses of future datasets. The second deployment described here primarily concerns first observations of the Crab Nebula. 

\section{Data and Results}
Observations of the Crab Nebula were conducted on the evenings of 30 October--1 November 2024 (UTC) at Lick Observatory, Mt Hamilton, California. Each telescope was installed, powered, and configured to the network in only a few days, which was made possible by taking advantage of existing infrastructure at the observatory. Weather conditions were not ideal. In some instances, high humidity caused water to condense on the Fresnel lenses. In the case of rain, telescopes and electronics were brought inside until observing conditions were more favorable. Telescopes were realigned and re-calibrated as needed.

Each telescope triggers when the intensity of light reaches a configurable threshold in a programmable number of pixels. By requiring that two pixels located on the same quarter of the camera (containing 256 pixels) be above threshold, we are able to reduce the frequency of false triggers due to night sky background. We typically operated with this two pixel-requirement at a threshold of 6.5 photoelectrons. During these nights, Jupiter was within a few degrees of the Crab Nebula, so we fixed the telescope pointings to RA 05h 44m 32s, Dec +22$^\circ$ 01$'$ 03$''$, to reduce Jupiter's influence on the observations. Telescope trigger thresholds were also raised to 7.5 photoelectrons to maintain a stable data rate at roughly 1\,Hz per telescope. 

\begin{figure}[htb]
    \centering
    \includegraphics[width=\linewidth]{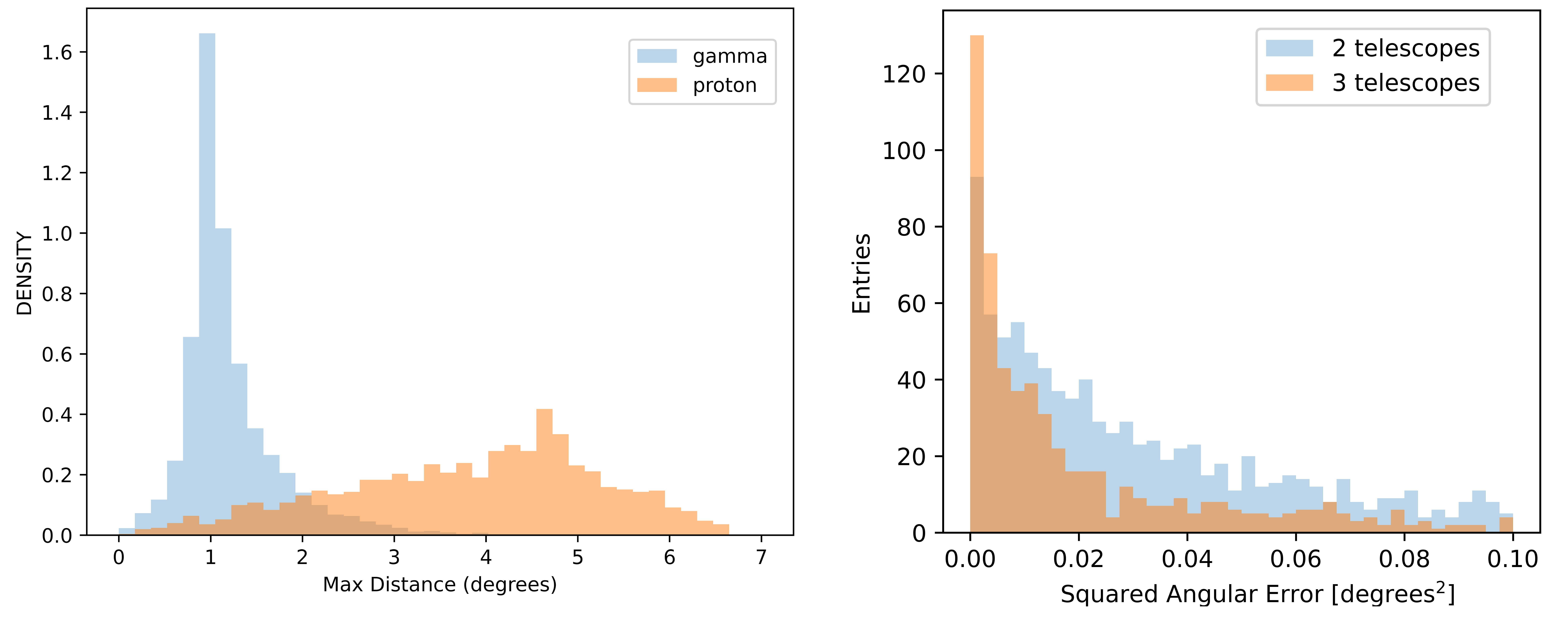}
    \caption{Histograms of the max distance parameter (left) and squared angular error of reconstructed arrival directions (right) for simulated data of the Lick Observatory array (see Figure \ref{fig:lick}). The max distance distributions are normalized to have integrated area equal to unity. The squared angular error, $\theta^2$, for reconstructed showers improves when more telescopes image the event in coincidence. When strictly two telescopes image the event, 68\% of reconstructed arrival directions fall within $0.4^\circ$ of the true simulated direction. When requiring 3 telescopes, the resolution improves to $0.19^\circ$. For the analysis of the Crab Nebula dataset presented here, we define simple cuts at max distance $< 2^\circ$ and $\theta^2 <0.32^\circ$.}
    \label{fig:cuts}
\end{figure}

We calculate Hillas parameters from cleaned images in order to distinguish gamma-ray showers from hadronic showers \cite{1985ICRC....3..445H}. Simulations of the array indicates strong separation between gammas and protons when using the \textit{distance} parameter, which measures the angular separation between the image centroid and some other test position in the camera plane (Figure \ref{fig:cuts}). To combine the information of the shower from all telescopes, we define a "max distance" parameter. This metric corresponds to the maximum value among distance parameters measured by all telescopes in a given event. The power of this cut is related to the geometry of the IACT technique, but is also largely a result of the PANOSETI telescope's wide field of view. The distance parameter is tightly correlated to the impact parameter of a gamma-ray initiated shower, and two competing effects shape its distribution. A shower that triggers a telescope is geometrically more likely to have impacted the ground at a large radial distance, but the visibility at this radial distance is limited by the intensity of the shower. These two effects produce a peak distance parameter between 1 and 2 degrees for gamma rays. Cosmic rays arrive isotropically, so there is no correlation with impact distance and position in the camera. They are simply more likely to appear on the edge of the camera where there is more area to view them. The tail in the proton distribution is due to the square nature of the camera, which constrains the field of view at the edges. 

We also reject events whose reconstructed arrival directions fall more than $0.32^\circ$ outside the test position. This corresponds to the 68\% containment radius for reconstructed simulations that triggered at least 2 telescopes. The histograms of squared angular error, $\theta^2$, for strictly 2, and strictly 3 telescopes are plotted for comparison in Figure \ref{fig:cuts}.  Both the max distance and $\theta^2$ cuts rely on a source position known a priori, which we deem sufficient for a point-like analysis of the Crab Nebula.

\begin{figure}[htb]
    \centering
    \includegraphics[width=\linewidth]{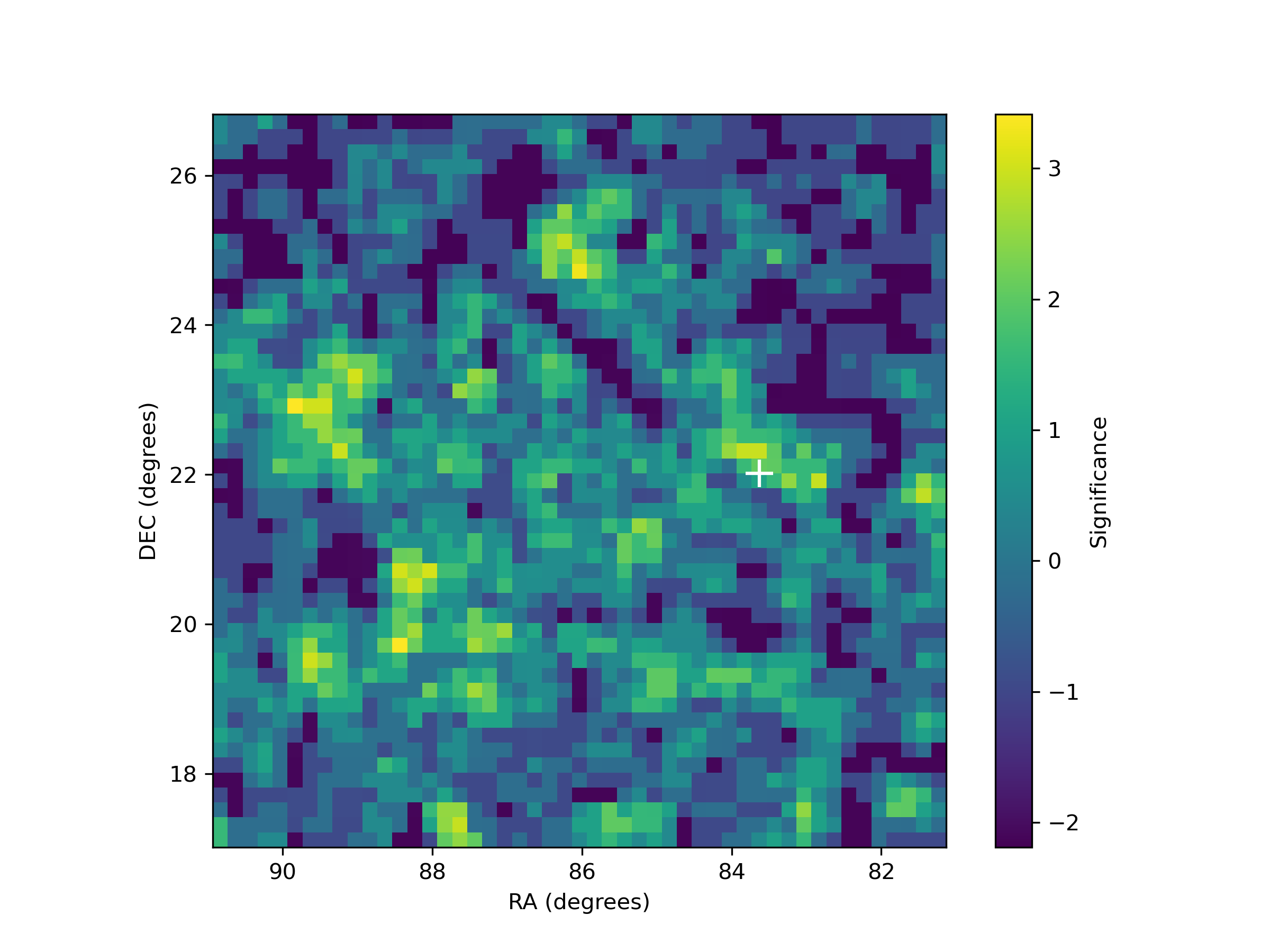}
    \caption{Significance map for the 5 hour Crab Nebula dataset. The location of the Crab nebula is marked by a white "+". The Crab was positioned outside the center of the field of view due to the close proximity of Jupiter on the nights these data were taken. The Li \& Ma significance at the position of the Crab Nebula corresponds to $1.55\,\sigma$.}
    \label{fig:sigmap}
\end{figure}

Significances are calculated according to Equation 17 of Li \& Ma \cite{1983ApJ...272..317L}. A map of significances across the full field of view is plotted in Figure \ref{fig:sigmap}. At the position of the Crab Nebula, we report a nondetection with a significance of $1.55\, \sigma$. We measured 5 on-source counts, and 248 off-source counts taken from most of the remaining field of view. The scaling factor is $\alpha=1/109$, yielding an excess of $N_\text{ON}-\alpha N_\text{OFF}=2.7$ events. This result is not a surprise when factoring in the poor quality of data, and only a 5 hour exposure with a $\sim10$\,TeV energy threshold. However, if we assume an optimistic effective area between $10^4$--$10^5$\,m$^2$ at the energy threshold, we only expect to measure a handful of excess counts in 5 hours, which is consistent with the Crab spectrum measured by LHAASO \cite{2021Sci...373..425L}. It is also consistent with the PANOSETI observations conducted jointly with VERITAS \cite{2022SPIE12184E..8BM}.

\section{Status of Deployment at Palomar Observatory}
\begin{figure}[htb]
    \centering
    \includegraphics[width=\linewidth]{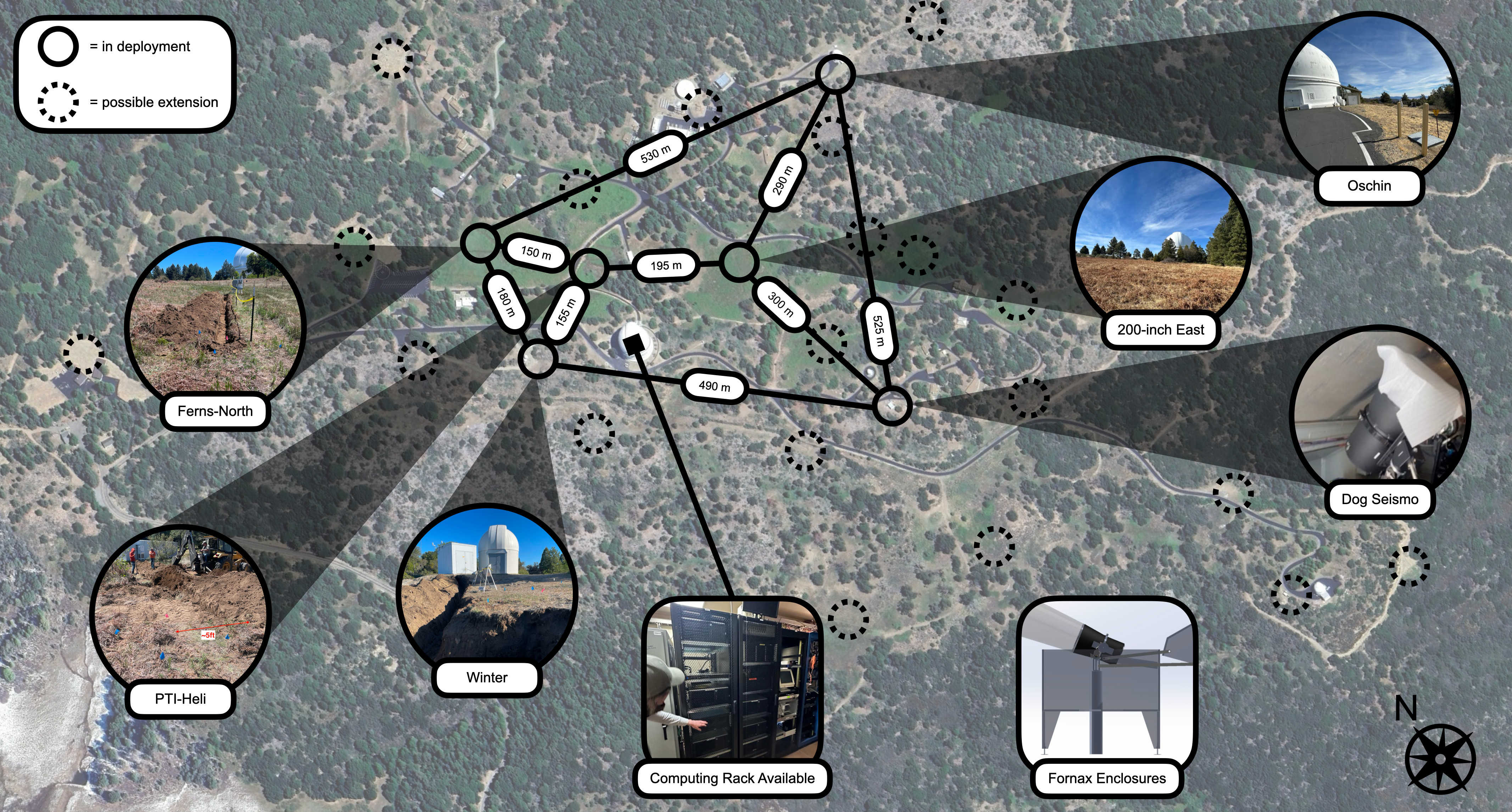}
    \caption{Layout of the 2024 Dark100 installation at Palomar Observatory. Cable trenching has already begun at the Ferns-North, PTI-Heli and Winter sites. A PANOSETI telescope is being deployed inside an enclosure already constructed at the Dog Seismo site. All other locations will be deployed on a metal pier cemented to the ground, and contained within a Fornax enclosure. A computing rack for Dark100 data acquisition is available in the basement of the Hale Telescope dome  Maps Data: Google, \copyright 2025 Airbus. }
    \label{fig:palomar}
\end{figure}

Dark100 is a new gamma-ray observatory currently under construction at Palomar Observatory (Figure \ref{fig:palomar}). Dark100 will search for ultra heavy dark matter particles ($m_\chi > 100$\,TeV) by observing UHE gamma-rays as a potential indirect signature. Dark100 will also study the UHE gamma-ray objects that have been detected on the Galactic Plane. The wide field of view of PANOSETI telescopes is particularly advantageous in this regard, and also facilitates monitoring of the Galactic Center. In the event that dark matter is not detected, Dark100 will still provide a new perspective in the search for Galactic PeVatrons.

The array is already funded to build up to 6 PANOSETI telescopes, with plans to build more in the future. Cable trenching has already begun at the Ferns-North, PTI-Heli, and Winter sites. These telescope will be deployed outdoors on metal piers cemented to the ground, within custom-designed Fornax\footnote{Fornax: \url{https://fornaxmounts.com/}} enclosures. A PANOSETI telescope will also be installed inside a building that already exists at the Dog Seismo site. The first three telescopes at Ferns-North, PTI-Heli, and Winter form a triangular footprint similar to the previous Lick Observatory array. These three telescopes will begin collecting data before the end of this year, and all the installed telescopes will remain at Palomar observatory to operate for at least 5 years.

\section{Summary}
With 5 hours of data on the Crab Nebula, we calculate a significance of $1.55\,\sigma$ consistent with background fluctuation. However, this result was obtained with three 0.5-m telescopes, each roughly 5\% the cost of a traditional IACT. Perhaps more remarkably, the whole array was deployed, operated and decommissioned in only one week. A more detailed model of the telescopes is being developed to more accurately measure instrument response, and to refine the cut selection to be robust for future observations. The analysis of the Crab Nebula presented here is a proof of concept that shows an array of sparsely separated Fresnel lens IACTs is a viable path toward a gamma-ray observatory optimized at UHE. This principle led to the conception of the successfully funded Dark100 project. Up to six telescopes are going to be deployed at Palomar observatory very soon, and the first three will begin observations before the end of this year. Dark100 is planned to operate for at least five years, and will provide insights on an underexplored region of dark matter parameter space in addition to providing a high resolution glimpse at Galactic PeVatrons. 

\section*{Acknowledgments}
We thank the Lick Observatory staff and engineers for their help in the installation of the PANOSETI telescopes at the Crocker, Dorm, and Kron sites. We also thank the Palomar Observatory staff and engineers for their help in the ongoing installation of the Dark100 array. The PANOSETI research and instrumentation program has been made possible by the enthusiastic support and interest of Franklin Antonio and the Bloomfield Family Foundation. Harvard SETI was supported by The Planetary Society. UC Berkeley’s SETI efforts involved with PANOSETI are supported by NSF grant 1407804, and the Marilyn \& Watson Alberts SETI Chair fund. E. Pueschel, Y. Popovych and S. Ravikularaman  are supported by an ERC Consolidator Grant (No. 101124914). The PANOSETI program at Palomar Observatory is supported by the California Institute of Technology. We acknowledge support of the University of Delaware General University Research Program.

\bibliographystyle{JHEP}
\bibliography{bib}

\clearpage

\section*{\centering All Authors and Affiliations}
\centering PANOSETI Collaboration

\begin{small}
    A. Brown\textsuperscript{1},
    B. Godfrey\textsuperscript{2},
    J. Holder\textsuperscript{3},
    P. Horowitz\textsuperscript{4},
    A. Howard\textsuperscript{5},
    A. Johnson\textsuperscript{1},
    N. Korzoun\textsuperscript{3},
    W. Liu\textsuperscript{2},\\
    J. Maire\textsuperscript{1},
    Y. Popovych\textsuperscript{6},
    E. Pueschel\textsuperscript{6},
    N. Rault-Wang\textsuperscript{2}
    S. Ravikularaman\textsuperscript{6},
    D. Werthimer\textsuperscript{2},
    S. A. Wright\textsuperscript{1}
    
\end{small}

\vspace{1em}
{\footnotesize
\noindent
\textsuperscript{1}Department of Astronomy \& Astrophysics, University of California San Diego, 9500 Gilman Dr, La Jolla, CA 92093, USA \\
\textsuperscript{2}Department of Astronomy, University of California Berkeley, 501 Campbell Hall, Berkeley, CA 94720 , USA \\
\textsuperscript{3}Department of Physics and Astronomy and the Bartol Research Institute, University of Delaware, Newark, DE 19716, USA \\
\textsuperscript{4}Department of Physics, Harvard University, 17 Oxford St, Cambridge, MA 02138, USA \\
\textsuperscript{5}Department of Astronomy, California Institute of Technology, 1200 E. California Blvd, Pasadena, CA 91125, USA \\
\textsuperscript{6}Astronomisches Institut, Ruhr-Universität Bochum, Universitätsstr. 150, 44801 Bochum, Germany \\
}

\end{document}